\documentclass[prd,onecolumn,showpacs,eqsecnum,amsfonts,amssymb,
amsmath,a4paper,english,superscriptaddress, nofootinbib
]{revtex4}
\usepackage{graphicx,multirow}
\usepackage[innercaption]{sidecap}

\begin{document}
\def\Barcelo{Barcel\'o}
\title{A real Lorentz-FitzGerald contraction}
\author{C. Barcel\'{o}}
\affiliation{Instituto de Astrof\'{i}sica de Andaluc\'{i}a, CSIC, Camino Bajo de Hu\'{e}tor 50, 18008 Granada, Spain}
\author{G. Jannes}
\affiliation{Instituto de Astrof\'{i}sica de Andaluc\'{i}a, CSIC, Camino Bajo de Hu\'{e}tor 50, 18008 Granada, Spain}
\begin{abstract}
Many condensed matter systems are such that their collective
excitations at low energies can be described by fields satisfying
equations of motion formally indistinguishable from those of
relativistic field theory. The finite speed of propagation of the
disturbances in the effective fields (in the simplest models, the speed
of sound) plays here the role of the speed of light in fundamental
physics. However, these apparently relativistic fields are immersed in
an external Newtonian world (the condensed matter system itself and the
laboratory can be considered Newtonian, since all the velocities involved
are much smaller than the velocity of light) which provides a
privileged coordinate system and therefore seems to destroy the possibility of having a
perfectly defined relativistic emergent world.
In this essay we ask
ourselves the following question: In a homogeneous condensed matter medium, is
there a way for internal observers, dealing exclusively with the
low-energy collective phenomena, to detect their state of uniform
motion with respect to the medium? By proposing a thought experiment
based on the construction of a Michelson-Morley interferometer made of
quasi-particles, we show that a real Lorentz-FitzGerald contraction
takes place, so that internal observers are unable to find out anything
about their `absolute' state of motion. Therefore, we also show that
an effective but perfectly defined relativistic world can emerge in a
fishbowl world situated inside a Newtonian (laboratory) system. This
leads us to reflect on the various levels of description in physics,
in particular regarding the quest towards a theory of quantum
gravity.

\end{abstract}
\maketitle
\section{Introduction: Emergent metrics in condensed matter systems}
It is by now a well known fact that the physics associated with
fields, classical or quantum, in curved backgrounds can be reproduced
in a large variety of condensed matter systems, the so-called analogue
models of general relativity~\cite{barcelo2005,novello2002}. The key
point is the realization that some collective properties of these
condensed matter systems satisfy equations of motion formally
equivalent to that of a relativistic field in a curved spacetime. The
simplest example is the equation describing a massless scalar field
$\phi$,
\begin{eqnarray}%
{1 \over \sqrt{-g}} \partial_\mu \sqrt{-g} g^{\mu\nu} \partial_\nu \phi =0,
\end{eqnarray}%
satisfied, for example, by acoustic perturbations in a moving perfect
fluid. These acoustic perturbations therefore travel along the
null geodesics of the acoustic metric $g_{\mu\nu}$. Generically, the most
interesting relativistic behaviours show up in the analysis of
collective excitations from the vacuum state of specific condensed
matter systems~\cite{volovik2003}. In those cases, one can say that relativistic effects
have emerged in the low-energy corner of the corresponding theory, since the relativistic symmetry was not present at higher energies nor at the microscopic level.

When looking at the analogue metrics one problem immediately comes to
mind. The laboratory in which the condensed matter system is set up
provides a privileged coordinate system.
Thus, one is not really reproducing a geometrical configuration but only a specific metrical representation of it. This naturally
raises the question whether diffeomorphism invariance is not lost in
the analogue construction. Indeed, if all the degrees of freedom
contained in the metric had a physical role, as opposed to what happens in
a general relativistic context in which only the geometrical 
degrees of freedom (metric modulo diffeomorphism gauge) are
physical, then diffeomorphism invariance would be violated.

The standard answer to this question is that diffeomorphism invariance
is maintained but only for internal observers, i.e.\ those observers
who can only perform experiments involving the propagation of the
relativistic collective fields. In this essay we want to elaborate on
this problem. Which are the necessary conditions in order for the
world description developed by such internal observers to be
completely relativistic?

Let us take the simplest geometrical configuration, Minkowski
spacetime. Imagine for example that we have a homogeneous fluid (or
more generally, a homogeneous medium) at rest in the laboratory. The
analogue metric associated with this configuration would be the Minkowski
metric. Is an internal observer capable of discerning whether he is at
rest in the medium or moving through it at a certain uniform velocity?

\section{The Michelson-Morley interferometer}

We were told at school that a way to distinguish whether one is moving
or not with respect to a medium is to set up a Michelson-Morley type
of experiment~\cite{michelson1887}. In our analogue version of this
experiment an interferometer will be used with two arms of perfectly
equal length which are placed perpendicularly and along which
``acoustic signals'' are sent (here we will designate as acoustic
signal any signal in the form of a relativistic massless field
perturbation; these signals travel at a constant speed, the speed of
sound $c_s$). Mirrors placed at the ends of these arms reflect the acoustic signals back to a common point $p$. If a displacement of the interference fringes at $p$ is observed when the interferometer is rotated, then one can conclude that the velocity along let's say the $x$ axis is different from that along the $y$ axis.

However, there is a hidden assumption in this experiment. Although the
internal observers seem to be using only acoustic signals, the
interferometer itself is an apparatus completely alien to the medium
and its excitations. By assuming the availability of this
interferometer for the usage of internal observers, one is also
assuming that they are in contact with the outside or external world
(anything not describable in terms of collective excitations within 
the system). A genuinely internal observer should be confined to the
manipulation of objects strictly within his own realm. Consistently,
internal observers can only use an interferometer if it could have
been created by themselves. So, imagine that instead of using an
interferometer in which the two rigid arm structures are built by putting
together particle after particle, they were using a
quasi-interferometer created by using \textit{quasi-particles} as building
blocks.

A quasi-particle is a collective excitation of the system with
properties similar to those of particles. Imagine that among the
emergent features of the condensed matter-like system there exist some
quasi-particles with mass so that they can be at rest within the
system. These massive quasi-particles could be genuine, as in a
two-component Bose--Einstein condensate~\cite{visser2005}, or complex
systems formed themselves by elementary quasi-particles. Imagine
also that these quasi-particles interact with each other through the
distortion of the acoustic fields, as is the case e.g.\ for quasi-neutrinos in $^3$He-A~\cite{volovik2003}\footnote{In this essay we will not discuss the feasibility of the combination of these two hypotheses in current realistic condensed matter systems; we just take the view that within a sufficiently complicated condensed matter system it should be possible to fulfil them both.}. 

We are going to show that these two hypotheses generically imply that the physical length of a quasi-interferometer
arm, as measured in the lab, would shrink by an acoustic Lorentz
factor $\gamma=\left({1-v^2/c_s^2}\right)^{-1/2}$ when moving at a velocity $v$ with respect to the medium ($c_s$ is the velocity of sound in
the medium).

\section{A real Lorentz-FitzGerald contraction}

Let us consider the following specific situation: We have {\it i)} 
a medium at rest, \mbox{{\it ii)} some} relativistic collective 
excitations described 
by a field $A_\mu$ and \mbox{{\it iii)} a} massive quasi-particle
acting as a source with charge $q$ of the field $A_\mu$, so 
that the equations satisfied by the relativistic field are
\begin{eqnarray}%
\Box A_\mu - \partial_\mu (\partial^\nu A_\nu)= j_\mu;
~~~~
j_\mu= \left\{
-q\delta^3[\vec x-\vec x(t)], q (\vec v /c_s) \delta^3[\vec x-\vec x(t)]  
\right\},
\end{eqnarray}%
with $\vec x(t)$ the trajectory of the massive quasi-particle and
$\vec v=d \vec x(t) /dt$ its velocity. The d'Alembertian corresponds
to that of a Minkowskian acoustic metric so that by using the three
Cartesian coordinates of the lab and the absolute laboratory time it
can be written in the standard Cartesian form. The field equations are
expressed in this way only when using these lab coordinates and when
the medium is at rest. Note that we are considering a vector field because of
its parallelism with the (electromagnetic) field responsible for the
existence of solid bodies (and so of rigid rods) in nature. However, many
of the following arguments could have been presented using a single
massless scalar field. Let us also point out that there exist specific
condensed matter systems (for example $^3$He-A~\cite{volovik2003}) in
which fields and field equations like these emerge in the low-energy
description of the system.

Let us first solve the system when the source is at rest ($\vec v=0$). A solution for $A_\mu$ is then
\begin{eqnarray}%
A_0 = -{q \over [(x-x_0)^2+(y-y_0)^2+(z-z_0)^2]^{1/2}}; ~~~~ A_i=0.
\end{eqnarray}%
There are other solutions to the previous equations, but these are
all related to one another by a gauge transformation
$A'_\mu=A_\mu+\partial_\mu \chi$.  For simplicity, let us choose the
Lorenz gauge $\partial^\mu A_\mu=0$ and continue with the discussion.

We now seek for a solution of the same system of equations
but with a source moving uniformly in the $x$ direction:
\begin{eqnarray}%
j_\mu= &&\hspace{-2mm}
\{-q\delta[x-(x_0+vt)]\delta(y-y_0)\delta(z-z_0),\nonumber
\\
&&q (v/c_s) \delta[x-(x_0+vt)]\delta(y-y_0)\delta(z-z_0),0,0 \}.
\end{eqnarray}%
The solution can be found in many text books (see for 
example~\cite{jackson1998}) and reads
\begin{eqnarray}%
&&A_0(x) = 
-{q \gamma \over [(\gamma(x-v t) - \gamma x_0)^2+(y-y_0)^2+(z-z_0)^2]^{1/2}}; \nonumber
\\
&&A_x(x)= {q \gamma (v/c_s)\over [(\gamma(x-v t) - \gamma x_0)^2
+(y-y_0)^2+(z-z_0)^2]^{1/2}};
\\
&&A_y(x)=A_z(x)=0.\nonumber
\end{eqnarray}%
The only difference with the standard solution for a proper
electromagnetic field is that in this solution the gamma factor is
defined with the velocity of sound $c_s$ rather than with the
velocity of light. The important point we want to highlight here is
that the fields decay faster in the $x$ direction than in the
orthogonal $y,z$ directions, the ratio of the two decays being given by
a sonic $\gamma$ factor.

The next step in our construction of a quasi-interferometer is to use
these emergent vector fields and sources to produce a rigid
bar. The simplest model one can think of for that task is provided by
dipolar interactions between globally neutral objects. Imagine
that it is possible to create, starting from elementary charged
sources (i.e.\ charged quasi-particles), a neutral super-structure. This super-structure
would correspond to  something that we could call a `quasi-atom'. The
existence of rigid structures in nature is based on the possibility
for atoms to be arranged in a regular and stable way. To make this
possible, two atoms must find their most stable configuration when
they are separated from each other a distance $a_0$. Here we
are going to reproduce with quasi-atoms in our condensed-matter model what is found in nature for atoms in solids.

Take two of these neutral objects (quasi-atoms) and place them at rest with
respect to the medium, both along the $x$ axis, and at a distance $a$
with respect to each other. If one associates an energy density $e$
with the vector field created by the total system, then this
energy density will be a function of the location in space and of the
distance $a$: $e=e(\vec x, a)$. Integrating over the entire space
one finds an interaction energy
\begin{eqnarray}%
E_I(a)=\int dx^3 e(\vec x,a).
\end{eqnarray}%
In suitable situations one will obtain that this interaction energy
has a sharp minimum at $a=a_0$, exactly what is needed to guarantee
the rigidity of the two-component system.

Now, assume that the two neutral objects are moving uniformly in the
\mbox{$x$ direction} while maintaining their relative distance. The energy
density will now depend on $a$, $\vec x$ and $t$. Again, after
integration one will find
\begin{eqnarray}%
E'_I(a)=\int dx^3 e'(\vec x,t,a).
\end{eqnarray}%
According to the result obtained for the one-particle case,
namely that the vector field associated with a moving charge acquires
a gamma factor in its decay in the $x$ direction compared to the case for the charge at rest, it is natural to expect that this interaction
energy will be precisely $E'_I(a)= E_I(\gamma a)$. The minimum of this
potential energy satisfies the condition
\begin{eqnarray}%
0={d E'_I(a) \over d a}= \gamma {d E_I(a') \over d a'}; ~~~~~~~~~ 
a'= \gamma a
\end{eqnarray}%
Thus, the minimum now occurs when $a'=a_0$, that is, when $a=
\gamma^{-1} a_0$. In this way we have shown that the real distance
(i.e.\ the distance measured in the lab) between the two quasi-atoms has
decreased by an acoustic Lorentz factor $\gamma$ due to their velocity
with respect to the medium. And obviously, when adding more quasi-particles to construct the quasi-interferometer arm, the Lorentz factor in the $x$ direction will persist.

This acoustic gamma factor is precisely the length contraction -- the contraction of an interferometer arm oriented in the direction of motion -- which is needed in order for the interference pattern to remain unaffected by such a uniform motion.
So a Michelson-Morley type of experiment using a
quasi-interferometer does not make it possible for internal observers
to distinguish whether they are at rest or uniformly moving with
respect to the medium. Therefore, a natural description of the
internal world (based on operational definitions) as experienced by
its internal inhabitants is provided by relativity theory. 

In this simple model, this relativistic world in a fishbowl is itself immersed in a
Newtonian external world (the laboratory). Remarkably, all of relativity (at least, all
of special relativity) could be taught as an effective theory by using
only Newtonian language.\footnote{A similar point of view was defended by Bell \cite{bell1976}.}

So we have explicitly demonstrated here for the case of a flat spacetime what was suggested earlier in, for example, \cite{liberati2002} and \cite{volovik2003}, namely that
Lorentz invariance is not broken, i.e., that an internal observer cannot detect his absolute state of motion. The argument is the same for curved spacetimes: The internal observer would have no way to detect the ``absolute'' or fixed background. So the apparent
background dependence provided by the (non-relativistic) condensed matter system will not violate diffeomorphism invariance, at least not for these internal inhabitants. These internal observers will then have no way to collect any metric information beyond what is coded into the intrinsic geometry (i.e., they only get metric information up to a gauge or diffeomorphism equivalence factor). Internal observers would be able to write down diffeomorphism invariant Lagrangians for relativistic matter fields in a curved geometry. The dynamics of this geometry, however, is a different issue. It is a well-known problem
that the expected relativistic dynamics, i.e. the Einstein equations, have not been reproduced in any known condensed matter system so far.

\section{Searching in the past}

In a way, the model we are discussing here could be seen as a
variant of the old ether model. At the end of the 19th century, the
ether assumption was so entrenched in the physical community that,
even in the light of the null result of the Michelson-Morley
experiment, nobody thought immediately about discarding it. Until the
acceptance of special relativity, the best candidate to explain this
null result was the Lorentz-FitzGerald contraction hypothesis. What
Lorentz and FitzGerald proposed was essentially what we have described
in the previous section, namely\footnote{FitzGerald published a similar
suggestion without formally working it out
\cite{fitzgerald1889}. Curiously, even FitzGerald himself was not
certain whether it had effectively been published or not and this was
only `discovered' in 1967
\cite{brush1967}.}:
\vspace*{2mm}

\begin{minipage}{0.9\columnwidth}
``one would have to imagine that
the motion of a solid body (...) through the resting ether exerts upon
the dimensions of that body an influence which varies according to the
orientation of the body with respect to the direction of
motion.''\cite{lorentz1895}
\end{minipage}
\vspace*{2mm}

There are important differences between our thought model and the idea
of a luminiferous ether as it was held at the end of the 19th
century. For example, in our model everything --- spacetime, electromagnetism
and \mbox{(quasi-)matter} --- arises effectively from the same condensed matter
`ether', whereas the luminiferous ether supposedly only affected
electromagnetic phenomena. In addition, we consider our model of a
relativistic world in a fishbowl, itself immersed in a Newtonian
external world, as a source of reflection, as a
\textit{Gedankenmodel}. By no means are we suggesting that there is a
world beyond our relativistic world describable in all its facets in
Newtonian terms. 

Coming back to the contraction hypothesis of Lorentz and FitzGerald, it is
generally considered to be \textit{ad hoc}. However, this
might have more to do with
the caution of the authors, who themselves presented it as a
hypothesis\footnote{Lorentz uses the term
\textit{H\"ulfshypothese}. Holton \cite{holton1969} has observed that
Lorentz introduces at least eleven hypotheses in
\cite{lorentz1904}. Although some of these seem much more natural than
others, Lorentz introduces all of them in the same cautious way: as
hypotheses that are needed to advance with the theory.}, than with the
naturalness or not of the assumption. The contraction hypothesis is in fact quite
natural if one assumes that not only electromagnetism but also matter
(and in our condensed matter model, even spacetime itself) are
effective phenomena emerging from the ether. As observed by
Lorentz\footnote{Or in FitzGerald's words:``We
know that electric forces are affected by the motion of the
electrified bodies relative to the ether, and it seems a not
improbable supposition that the molecular forces are affected by the
motion, and that the size of a body alters consequently''
\cite{fitzgerald1889}.}
:

\vspace*{2mm}
\begin{minipage}{0.9\columnwidth}
``Surprising though this hypothesis may appear at first sight,
yet we shall have to admit that it is by no means far-fetched, as soon
as we assume that molecular forces are also transmitted through the
ether, like the electrical and magnetic forces. [Then] the translation
will very probably affect the action between two molecules or atoms in
a manner resembling the attraction or repulsion between charged
particles.''\cite{lorentz1895}
\end{minipage}
\vspace*{2mm}

The reason that special relativity was considered a better explanation
than the Lorentz-FitzGerald hypothesis can best be illustrated by
Einstein's own words: 

\vspace*{2mm}
\begin{minipage}{0.9\columnwidth}
``The introduction of a `luminiferous ether'
will prove to be superfluous inasmuch as the view here to be developed
will not require an `absolutely stationary space' provided with
special properties.''\cite{einstein1905}
\end{minipage}
\vspace*{2mm}

The ether theory had not
been \textit{disproved}, it merely became
\textit{superfluous}. Einstein realised that the knowledge of the
elementary interactions of matter was not advanced enough to make any
claim about the relation between the constitution of matter (the
`molecular forces'), and a deeper layer of description (the `ether')
with certainty. Thus his formulation of special relativity was an advance within the
given context, precisely because it avoided making any claim about the
fundamental structure of matter, and limited itself to an
\textit{effective} macroscopic description. That this was very clearly
realised by Einstein himself can be seen from the following
quote. 

\vspace*{2mm}
\begin{minipage}{0.9\columnwidth}
``The next position which it was possible to take up in face of
this state of things [the acceptance of the special theory of
relativity] appeared to be the following. The ether does not exist at
all. (...) More careful reflection teaches us, however, that the
special theory of relativity does not compel us to deny ether. We may
assume the existence of an ether; only we must give up ascribing a
definite state of motion to it.''
\end{minipage}
\vspace*{2mm}
\newline
And he continues with the following model:

\vspace*{2mm}
\begin{minipage}{0.9\columnwidth}
``Think of waves on the surface of water. Here, we can describe
two entirely different things. Either we may observe how the
undulatory surface forming the boundary between water and air alters
in the course of time; or else -- with the help of small floats, for
instance -- we can observe how the position of the separate particles
of water alters in the course of time. If the existence of such floats
for tracking the motion of the particles of a fluid were a fundamental
impossibility in physics -- if, in fact, nothing else whatever were
observable than the shape of the space occupied by the water as it
varies in time, we should have no ground for the assumption that water
consists of movable particles. But all the same we could characterise
it as a medium.''\cite{einstein1920}
\end{minipage}
\vspace*{2mm}

\section{Some final remarks}

In Einstein's pragmatic approach for explaining the null result of the
Michelson-Morley experiment\footnote{Although historically, when
formulating special relativity, Einstein was barely aware of, let alone concerned with, the Michelson-Morley experiment in particular \cite{holton1969}.}, and in
particular in his analogy of the waves on the surface of water
described in the previous paragraph, lie the essential lessons that we
can learn from the fishbowl model presented in this essay. As long
as we have no possibility to directly track the elementary level of
description that we are looking for in a quantum theory of
gravity (the particles that constitute the water, in Einstein's
analogy), maybe we should limit ourselves to a description of how
spacetime (the undulatory surface at the boundary between water and
air) and matter effectively emerge from this elementary level (the
ether or medium), without making any more assumptions about the water
particles themselves than strictly necessary.

The thought model that we have presented shows that, as long as there
are no direct experimental constraints on this elementary description,
one should take care when postulating which aspects of the currently
known `effective' physics (i.e.\ of the internal world) should also be
taken as fundamental in the elementary description (the external
world). As a matter of fact, in our \textit{Gedankenmodel}, we
showed that this fundamental external world can even be Newtonian (the
laboratory), while still reproducing a relativistic behaviour with
respect to an internal observer. Only if this internal observer were able
to probe the external world, e.g.\ through observation of
phenomena linked directly to a quantum regime of gravity, would he be
able to really know anything about the fundamental physics. Of course,
it is far from certain that nature will actually allow us to probe
this deeper layer of description, i.e.\ maybe 

\vspace*{2mm}
\begin{minipage}{0.9\columnwidth}
``tracking the
motion of the particles on a fluid [is] a fundamental impossibility in
physics.''\cite{einstein1920}
\end{minipage}
\vspace*{2mm}
\newline
But even if this were the case, this
would not necessarily mean the end of the story. If it turns out that
nature's defences are strong enough to prevent us from probing the
quantum scale of gravity, then maybe this intrinsic protection should
precisely be used as a new theoretical guideline.

So when Einstein concluded in \cite{einstein1920} that

\vspace*{2mm}
\begin{minipage}{0.9\columnwidth}
``according to
the general theory of relativity space without ether is
unthinkable. (...) But this ether should not be thought of as endowed
with the quality characteristic of ponderable media, as consisting of
parts which may be tracked through time. \textit{The idea of motion
may not be applied to it},''
\end{minipage}
\vspace*{2mm}
\newline
we would add that the idea of motion may
not be applied to it in the same sense as it is for an internal
observer. As long as the fundamental physics remains hidden from
direct observation by internal observers, there could be 
\textit{plenty of states of motion that may be applied to it}. 

\section*{Acknowledgements}

The authors wish to thank Andr\'es Cano and Luis J. Garay 
for very useful comments and suggestions. This work has been funded by the Spanish MEC under project FIS2005-05736-C03-01. G.J. was also
supported by CSIC grant I3P-BPD2005 of the I3P
programme, cofinanced by the European Social Fund.



\end{document}